# Ethical Underpinnings in the Design and Management of ICT Projects

Aaditeshwar Seth
IIT Delhi and Gram Vaani, India
aseth@cse.iitd.ac.in, aseth@gramvaani.org

***Abstract*—With a view towards understanding why undesirable outcomes often arise in ICT projects, we draw attention to three aspects in this essay. First, we present several examples to show that incorporating an ethical framework in the design of an ICT system is not sufficient in itself, and that ethics need to guide the deployment and ongoing management of the projects as well. We present a framework that brings together the objectives, design, and deployment management of ICT projects as being shaped by a common underlying ethical system. Second, we argue that power-based equality should be incorporated as a key underlying ethical value in ICT projects, to ensure that the project does not reinforce inequalities in power relationships between the actors directly or indirectly associated with the project. We present a method to model ICT projects to make legible its influence on the power relationships between various actors in the ecosystem. Third, we discuss that the ethical values underlying any ICT project ultimately need to be upheld by the project teams, where certain factors like political ideologies or dispersed teams may affect the rigour with which these ethical values are followed. These three aspects of having an ethical underpinning to the design and management of ICT projects, the need for having a power-based equality principle for ICT projects, and the importance of socialization of the project teams, needs increasing attention in today's age of ICT platforms where millions and billions of users interact on the same platform but which are managed by only a few people.**

## I. INTRODUCTION

The optimism behind ICT projects being able to make the world a better place has visibly suffered a setback in recent times. In this essay, we try to answer the question of why this might be so, and suggest some conceptualization frameworks that can help build guidelines for ICT project designers and managers to ensure that responsible outcomes arise from ICTs.

The scepticism about the reliability of ICT projects to lead to positive outcomes is shared between both ICT4D and non-ICT4D projects. ICT4D projects, like most other development programmes, often start with a theory of change that will lead to certain development outcomes, then use a human-centred design approach to design the ICT elements, and finally deploy and iterate on the design through a series of pilots and scale-up phases. Non-ICT4D projects, that may be defined as those not conceived to primarily achieve some development objectives through a pre-determined theory of change, are not very different, and typically follow the same process of going from some objectives (even if not development oriented) to design and then to the deployment of these projects. Given the similarity in how these different types of projects are conceptualized and executed, the reasons for failure must be common too for both ICT4D and non-ICT4D projects, and the arguments in this essay may therefore be generalized to both.

Even with a high degree of forethought in defining the objectives and design of ICT projects, surprises however often seem to spring up during the deployment of these projects. For example, flexibilities designed into the ICTs for ease of use may could lead to misuse of the technologies and cause harm (eg. Facebook [1]), or inequalities in access to the technologies may manifest in skewed development outcomes (eg. digital gender divide [2]), or the technology selection may not be suited to the deployment context (eg. Aadhaar [3]). Many such problems also manifest slowly over time but the sooner they are identified and addressed, the better, because once the projects are scaled-up it becomes harder to change them, often due to cost considerations and vested interests that emerge for the continuation of the projects [4]. Methods like co-design and participatory design advocate for adequate pilot iterations and evaluation under diverse conditions so that such problems are recognized and strategies are developed to fix them before scaling the projects [5]. Since observations about the effects of the ICTs on development outcomes need a long-term evaluation though, and business or political imperatives may not favour slow and steady approaches, therefore such methods are typically unable to bring about strategic changes in most government and market-led projects. Methods like value-sensitive design take a pro-active approach by building certain well-defined values into the design itself so that chances for misuse or undesirable outcomes is minimized [6]. However, such methods may also create a false illusion of safety by design, by not emphasizing on the importance of value-sensitivity in the management of the deployment as well, ie. the need to deal with problems that will still arise despite extensive galvanization attempts made during the design phase.

Further, while rich literature exists for designing ICTs, such as [7, 8, 9], there is a paucity of studies about the management of ICT projects in terms of problems that arise during deployment and how to address them. Rich literature about deployment experiences does exist, such as [10, 11, 12], but it is mostly descriptive in terms of identifying the problems, not in terms of methodological approaches to find solutions to address the problems.

We make three arguments in this essay. First, we argue that managing the deployment of ICT projects deserves as much importance as their design, and that design alone cannot guarantee flawless deployment. Towards this, we propose a three-layer framework within which ICT projects can be conceptualized, starting with defining the objectives, then the design elements, and finally the deployment management strategies, with a clearly specified common ethical system underpinning all these three layers. The ethical system serves as a glue spanning all the layers, to resolve unforeseen problems or make choices or deal with uncertainty, which are likely to arise in practical situations right from framing the objectives to defining the design and building operating processes for managing deployments. The common ethical system brings consistency in resolving questions that might arise at any of the three layers.

Second, we identify several common patterns that lead to undesirable outcomes during the deployment phase. All these patterns interestingly seem to stem from how the power dynamics between actors involved in the ICT projects change as a result of introduction of the ICTs, and lead to creating new power inequalities or exacerbate existing ones. We suggest a framework to model these relationships both at the design stage to avoid designing projects that could lead to power differentials, and at the deployment stage to manage the operations so that power differentials are attentively neutralized before they can be misused. Put together with the first framework, we essentially suggest that power-based equality should be a key ethical principle that should shape the objectives, the design, and the management of ICT projects.

Three, we draw the focus towards the people who ultimately are the designers and managers of ICT projects, as the target audience for using the frameworks we have suggested here. These individuals who might own or conceive or design or operate ICT projects, are the ones who should critically examine their projects on the ethical values that underpin the objectives, design, and management of their projects. This becomes especially important in today's winner-takes-all ICT-driven platform era because the number of such people of responsibility is very few, but who ultimately end up shaping development outcomes for millions and billions of people. Understanding the ethical systems of these individuals, the processes of socialization between those who play different roles, and shaping of their ideologies by the wider political economy in which they operate, therefore becomes critical. Further, many such ICT projects actually claim to be addressing development outcomes such as empowerment (Facebook) or inclusion (Aadhaar), and therefore applying the ethics framework to such projects can reveal inconsistencies in claims made by the people running these projects.

To summarize, we try to draw attention to the fact that the design of ICT projects alone cannot guarantee flawless deployment, that the design and deployment of these projects is ultimately done by people, in today's winner-takes-all platform era this also means that the responsibly of running many huge ICT projects may rest in only a few hands, the ICT project teams actively make choices be it on the design or deployment or framing the objectives of their projects and therefore they need to operate through explicitly stated ethical principles that would shape the design and deployment process, and that aiming for an equitable power distribution needs to be a key ethical principle that the project teams should espouse. We do not prescribe a recipe to ensure responsible outcomes from ICT projects, rather we provide a lens through which ICT projects should be examined to keep a check on whether they are leading to positive outcomes or not.

## II. Ethical Underpinnings to ICT Projects

Figure 1 shows a three-layer structure that we propose to conceptualize ICT projects. The foundation of a project rests on its objectives. The design is influenced by the project objectives, the properties of the ecosystem within which the project is to operate, and any constraints or flexibilities that should to be imposed to manage the deployment. The deployment is influenced by the design in terms of the constraints and flexibility made available to manage the project. A common ethical system constituted of different values further influences all the three layers, and shapes the decision making process to define the objectives, the design, and the deployment processes. We explain this through three examples: A voice-based community media platform called Mobile Vaani (MV) operating in rural central India [10], the online social networking platform Facebook, and the unique identifier platform in India called Aadhaar.

### A. *Example: Mobile Vaani*

MV is founded on principles drawn from information science that when people from the same community share information it improves the contextualization of the message under discussion because of homophily effects, since people from the same community are likely to share the same context, and hence information shared by them helps others understand the message better [13]. On the other hand, diversity of viewpoints among the community members, especially contributed by those who hold bridging ties with other communities, helps improve the completeness of information and brings new insights [14]. This joint increase in context and completeness leads to people being able to understand the messages more quickly, discover new information, and counter their biases [15]. This in turn leads to faster utilization of the information, which could include actionable steps taken by people for their livelihood improvement, or drawing the attention of government officials to issues faced by marginalized groups, or shaping of social norms in related to health and gender practices [16]. With a focus on poor and less-literate populations, the objective of MV can therefore be specified as enabling a community media platform for information sharing that promotes greater context and completeness in the messages it carries [17, 18].

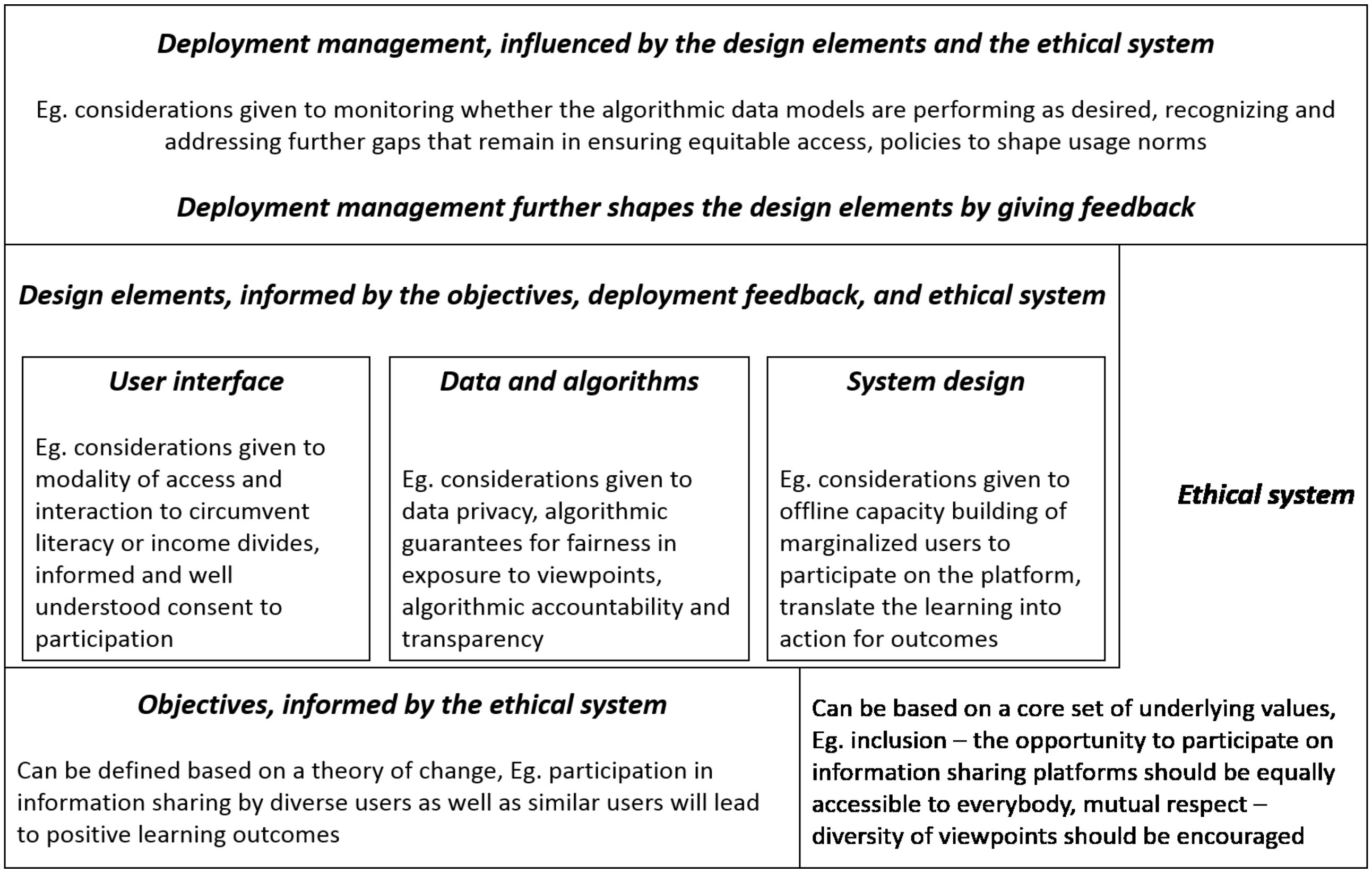

Figure 1: Underlying ethical conceptualization framework for ICT projects

This objective, and the ecosystem characteristics of less-literate and low-income communities in which MV is deployed, shaped its design from the outset. For example, the choice to use voice-based interaction, and content operation processes to create discussion among the users for context and completeness enhancement, were user interface and user experience related design decisions that were shaped by the underlying objectives and ecosystem characteristics. A strong focus on offline mobilization and training processes, with an attempt to bring participation beyond the IT-savvy and well-off rural populations, was incorporated as a design element specifically to create diversity and ensure equality in access to the platform [19]. Identification and training of community volunteers was undertaken to ensure embeddedness of MV into the community, so that it can carry messages and develop use-cases that are locally relevant to the communities. In the same way, moderation of content by manual review of messages recorded by the people, is done to ensure that a basic level of quality is maintained, and that norms are shaped towards voicing opinions but in a respectful tone and with an acknowledgement of diversity among people.

Unlike some of the above design elements that were conceived upon the inception of MV, new design elements also arose through observations during the deployment followed by a conscious effort to find solutions that could be baked into the design itself [20]. This includes some of the following. The stronger association that volunteers would place with their own class or caste or gender affiliations than with others, led to a concerted push towards creating diversity among the volunteer-base which could translate into creating diversity among the user-base. Further, a careful design of individual and group based incentives among the volunteers led to strong solidarity and mutual respect between them, along with eliminating free rider problems that often arise in collectives. Similarly, the need to establish norms for content contributions to create a conducive atmosphere for debate on the platform, led to evolution of a liberal yet disciplined and transparent editorial policy. This has so far been implemented through a central moderation team, and efforts are underway to decentralize the moderation processes to the volunteers themselves who are closely embedded in their communities. We therefore consider these design elements as requirements that arose from observations in the deployment, and were then incorporated into the MV design template itself.

The point we want to highlight is that the two sets of design elements listed above could very have been missed had the deployment not given them due attention [20]. The fact that the MV team during its early days was small and there was a high degree of socialization between the team members, certainly facilitated this deployment-design iteration step, but the socialization alone would not have led to incorporation of these elements into the design template. We argue that the incorporation became organizationally mandated due to the ethical system within which the MV team was operating. This ethical system was based on values of inclusion and equality so that marginalized groups would not get left out, values of respect to humans to include their voices even if their viewpoints were different, values of mentoring to guide them in using the platform especially to voice themselves, values of fairness to disallow free-riders among volunteers, and values of responsibility as platform providers to build a space conducive to debate and learning in line with the objectives of MV. This ethical system was responsible to notice the observations in the first place, then mobilize the iteration from deployment to design incorporation, and today it places emphasis in the deployment processes on ensuring adherence to the design through methods to monitor the activities [20].

The same ethical system emphasizing on values of inclusion, equality, and mentorship was also responsible for shaping the design elements conceived upon inception of MV, such as the use of voice to reach less-literate populations [21], the need for an offline training model to create technology literacy especially among less technology savvy users [19], and volunteer-led support for programmes like grievance redressal on issues with government schemes and services [22]. Even the platform objectives to create learning through conversation, can be argued as emerging from the same ethical principles that place value on each human and their thoughts [23]. Having a common ethical system therefore brought about consistency in how decisions were made at the different layers of the framework, and the socialization processes facilitated as a consequence of having a small team during the initial stages of the platform, were useful to build and operate MV in a responsible manner.

### B. *Example: Facebook*

Facebook has lately had to deal with several allegations about its working model, such as misuse of the platform to spread misinformation [24], and security loopholes that led to leakage of private user data [25]. We focus here on the former and use the framework to reason what could be the objectives, design decisions, deployment oversight, and the ethical system in which the Facebook teams would have been operating, that led to such misuse of the platform.

First, there seem to have been failures either in spotting misuse of the platform as part of deployment management, or in conveying these observations to iterate on the design so that new features or processes could be built, or in prioritization between addressing these observation in comparison with other deliverables. News articles point to lapses on potentially all three fronts [26]. However, the decision making process be it for prioritization or what to report versus what to ignore, is shaped ultimately by the underlying ethical systems in which the Facebook teams would have been operating. It points to missing values such as not creating a social contract on the platform that is founded on mutual respect by the users for one another, and values of responsibility felt by the platform providers to ensure a safe space for their users for online interactions on the platform.

Second, the solution adopted by Facebook to address these problems has been through a centralized review process that is partly algorithm driven and partly people driven, to adhere to community standards laid down by Facebook itself [27]. The community standards are arguably shaped by cultural norms of different communities and then codified as rules that can be implemented by humans and potentially even by machines, but such formalization of the inherent diversity of human society can clearly lead to misrepresentation and inaccuracy. The centralization of this function however to shape usage norms on the platform points to an inherent distrust in empowering the users to evolve norms themselves. So far there have not been any efforts by Facebook to reach out to communities using the platform and to train or mentor them, or provide technology features, through which the community managers can play a role to ensure responsible usage of the platform by their own members. This can be contrasted with Reddit's approach to provide the necessary tools and mentoring to managers of different subreddits [28], and again points to values such as respect for people to make decisions about what features of constraints or flexibilities should be provided pertaining to the usage of the platform.

Third, the argument above about a seeming distrust of the users, puts into question the stated objectives of Facebook to "give people the power to build community and bring the world closer" [29]. The stated objectives of ICT platforms therefore need to be examined critically against other observations made from the design and deployment layers, to evaluate for consistency in having a common ethical system across the layers. Further, since decisions about how to define the objectives or manage the deployments or iterate on the design, are ultimately made by people building and managing the ICT platform, this ethical system is nothing other than the values of these people. Further, the socialization processes between the people become an important determinant of which values become dominant or are heard.

### C. *Example: Aadhaar*

We next apply the same analysis method to evaluate the Aadhar unique identity system. A key stated goal of Aadhaar is to eliminate invalid identities such as when the same person may be in possession of multiple IDs and can use them to get double benefits [30]. The nationally valid identity provided by Aadhaar is also expected to support marginalized groups to avail welfare schemes from which they previously may have been excluded for not having adequate identity documents. Having a digital identity system operated through lowest common denominator technologies like biometrics that can be accessed by anybody, and linked with bank accounts for seamless payments and benefit transfers, was also expected to lead to painless technology adoption. Arguments have been raised about the rationality behind choosing these objectives. This includes concerns about whether problems of leakage of benefits can be solved by a better identity system, or does it rather require improved supply-chain tracking, and empowerment of the communities to demand accountability themselves [31]. Similarly to what extent is identity a barrier to availing benefits as compared to other documentation requirements, is questionable as well. We do not focus on the choice of objectives here, and restrict our analysis to exploring consistency in the ethical system underlying the chosen objectives, design, and deployment aspects of Aadhaar.

The stated goals clearly point towards values of fairness in the distribution of benefits, values of equality in accessing benefits, and even consideration of aspects of ease of use to remove capability or usability barriers in the technology adoption. However, these values do not reflect consistently across the layers. First, extensive reports have surfaced of how the technology design is not suitable for challenging rural conditions, which are marked by poor Internet connectivity, and even biometric matching errors of false negatives. This has led to exclusion of poor people from benefits, but the deployment observations have not strongly made their way to design changes, raising doubts on the commitment of the Aadhaar teams towards values of equality [3]. A similar gap is noticed with the inherent self-service design of the system with no room formally left for assisted usage. Deployment experiences point towards the need for less educated and poor people to take help from social workers and officials to rectify mistakes in the Aadhaar data [32], but this is done informally which in fact leads to security lapses, no changes in the design have however been incorporated to formalize some kind of assisted usage by trusted intermediaries. This again puts into question the values of the Aadhaar designers towards mentoring and easing usage for those people unable to deal with the technology, as opposed to expecting them to improve their skills or overcome other capability barriers to effectively use the system.

Second, no reports have so far been released about the efficacy of the biometrics to successfully eliminate duplicate identities. This raises doubts on fundamental values of honestly in whether the stated objectives were the real reason behind the choice of a biometrics based design of the platform, or was it unsaid objectives like building a national biometric database for security reasons.

Third, the centralized and non-transparent machine-driven decision making architecture with no easy appeal procedures for mistaken decision making, has ended up putting power back in the hands of the service providers who use Aadhaar to authenticate their transactions, and has taken power away from the hands of the consumers. This is seen extensively in the use of Aadhaar for authenticating PDS (Public Distribution System for subsidized food items) transactions. The PDS shop owner is able to leverage technology failure as a means to exercise power in different ways, eg. to deny ration, or to grant ration as a special favour rather than an entitlement, etc [3]. This increases the power differential between the shop owner and consumers even more, and can manifest itself in other spheres of community life where local elite may exploit the less powerful. In the same way, subservience to a centralized decision making system run by the government, is disempowering for the people because of the dependency and inescapability it creates for them towards a system that ultimately controls access to their life-critical entitlements. This fundamental design does not encourage power equality.

### *D. Detailed description of the framework*

Through the three examples we showed that the ICT project conceptualization framework can be used to analyse the objectives, design elements, deployment management processes, and relationships between these three layers, in terms of a common ethical system. It can reveal inconsistencies between the different relationships that may betray the stated claims of an ICT project. All the three projects analysed above were actually platforms that support ecosystems of many users and enable transactions between them, but we feel that the arguments can be generalized across non-platform ICT projects as well. In this section, we go into details of Figure 1 to propose the full-fledged framework that can accommodate the three examples.

The foundational layer of objectives is perhaps most easily articulated in terms of a theory of change of the project. In the case of MV, this theory of change is based on how information sharing will lead to learning, and eventually to individual and collective action based on the learning. With Aadhaar, this theory of change is based on how a unique digital identity will reduce corruption, improve entitlements, and ease access. The theory of change for Facebook is perhaps not as clearly articulated, it vaguely assumes that ease of online social communication will build communities but leaves the underlying processes undefined.

The objectives and assumptions about the theory of change lead to building relevant design elements at the inception stage of the ICT project. We have divided the design layer into three parts where we have seen significant research interest. First is the user interface where methods to enable the desired type of interactions and transactions between users, choices towards ease of use, the communication modality based on the literacy levels of users, procuring informed consent, generating persuasion through technology, etc, are the key design elements to configure. Second is the data and algorithms embedded in the technology, where design decisions about the privacy of data, fairness of algorithmic operations, access to accountability and appeal channels, etc, are some of the important design elements to choose. Third is a less researched area, of looking at the ICT project as being part of a larger ecosystem of actors, and the design being influenced by changes it may bring about in the relationships between these actors. The actors may have pre-existing relationships with one another, of power or cooperation or other dynamics between them, and the objective of an ICT4D project may in fact be to alter these dynamics. We discuss more about the part of system design later in the essay.

Design choices made in the examples can be mapped to these three parts in the design layer. For example, the choice of voice for MV to reach the less-literate userbase, or the choice of biometrics for Aadhaar, were made for the user interface based on the demographic characteristics of the intended users of these platforms and how best could access be enabled for them. The degree of automation for Aadhaar transactions, or testing for the compliance of messages shared on Facebook based on the community standards, were design choices made in the handling of data and algorithms at the design layer. The centralized architecture of Aadhaar and Facebook, or the decentralized volunteer driven appropriation of MV, were broader system design choices that were made when building these platforms.

The choice of design elements influences the actual usage during deployment by shaping norms of how users interact on the platforms, whether or not they are able to access the platforms as originally envisioned, whether the ICTs alter existing power relationships between different types of users, etc. A key function of deployment management is to astutely monitor the projects to assess whether the envisioned objectives are being realized, ie. whether the process of having encoded the objectives into the design, and made design choices with a view of influencing the usage, is actually working as desired. The deployment observations feed back into the design through an iterative process to alter the design so that usage during deployment falls in line with the desired goals. Incorporating deployment observations in MV's design to bring diversity in the user-base by creating diversity among the volunteer-base, or ignoring to re-design the Aadhaar technology despite deployment observations of technology failures, are examples of successful and broken links respectively between the deployment and design layers [20].

Finally, the ethical system directly influences each of the three layers in terms of stating the objectives, the design choices based on the objectives, the deployment management based on the design, and facilitating design iterations based on deployment observations. Ethical systems are constituted of values that the project teams honour, and could include values such as equality, inclusion, fairness, diversity, mentorship, mutual respect, honesty, responsibility, etc. For platform-based ICT projects, some values such as responsibility to create a conducive environment on the platform to realize the objectives, can be argued that they should be a part of the design itself, so that reacting to deployment observations becomes a fundamental expectation in the project rather than something that gets imposed through value-based prioritization of the project teams.

Our framing of the ethical system in terms of values is not meant to restrict the choice of the operationalizing ethical system to only virtue based ethics. Other ethical systems could serve as an underlying common mechanism as well. The government's attitude towards Aadhaar, for example, seems to be arising from a utilitarian approach [33] of looking at broader social good than pockets of failures. The Mobile Vaani approach of respecting human thought seems to be grounded in Kantian principles [33]. Our key message is that the choices made by project teams in ICT projects for the objectives, design, and deployment management, arise from inherent ethical systems in which the teams operate, and can hence be analysed with that viewpoint.

Note that our incorporation of values into the framework is different from how values are incorporated in value-sensitive design. Values in VSD are assumed to be baked into the design of a system and can serve as behaviour or usage bounding mechanisms to prevent misuse, by the users or by the designers of the system [6]. Values in our framework are the values espoused by the project teams that go into defining the objectives, design choices, acting upon the deployment observations, etc. These values are inherently subject to socialization biases and dynamics of the political economy that would shape the ICT project over time as a living artefact rather than something that can be designed once and will function according to the design specification henceforth. We discuss these socialization dynamics later in the paper. We next present an argument for a universal value that all ICT teams should incorporate, of power-based equality, and propose a method to incorporate this value as part of system design in the design layer. Before that however, we briefly present some related work to emphasize on the importance of an ethical grounding to technology design, and the less explored area of technology deployment that deserves just as much attention.

### *E. Related work*

#### *1) Ethics in design, and the insufficiency of achieving ethics by design*

The grounding of technology design in ethics has been pointed out often. [34] recognizes the inequality between designers and users, and that designers have a responsibility to ensure that their innovations create a just world and do good. This becomes challenging because designers may not always know their users, and hence it is suggested that they should operate using the Rawlsian principle of the veil of ignorance [33], so that biases of the designers do not hurt the users, especially the worst off users. It is further emphasized that principles of liberty and equality should form the foundations of technology design. Our work is on similar lines in emphasizing on the ethical foundations of technology design, and the responsibility of the designers themselves, but we outline a more detailed design framework which additionally distinguishes between design and deployment management, and we stress that ethical principles need to form a foundation for deployment management as well.

We also do not prescribe a particular form of ethical system suitable for ICT projects, and acknowledge Sen's criticism of the Rawlsian framework in its limitation that the veil of ignorance will also blind the designers to the current state of the users [35]. Therefore, equality will be hard to achieve as an eventual goal since the worst off users may not get access to the resources that can help them catch up to other users. Politics may indeed govern the choice of the ethical system that can be expected to achieve the desired political objectives.

This brings us to the realization that technology design can be shaped by the underlying politics of the designers, and that technology artifacts can favour one kind of political view over another, as explained by [36]. It is argued that technologies can build specific kinds of social order especially when wielded by powerful agencies such as authoritarian governments or even corporations and the media that may be driven by certain dominant ideologies. Under such circumstances, the technology design therefore may codify these ideologies and enforce alignment of entire societies towards these views. This is in agreement with our views that the underlying ethical systems in which the technology design teams would be operating, does get expressed in the choices they make in defining the objectives and design of the technology. However this view does not consider the potential of deployment feedback to shape design, when the designers and managers of the technology may be willing to adapt and learn to ensure responsible outcomes from their innovations.

As argued earlier, methods like VSD also restrict themselves to only emphasising that values should be embedded into the design itself and articulated explicitly, while we go further to say that values shape choices during the deployment as well, and that deployment feedback should continually shape the design. Another line of reasoning is to examine VSD as used in ICT platforms. VSD values are essentially context-free, such as data privacy or trust in technology, but ICT platforms may be used in widely different contexts which may demand dynamic adaptations in different situations based on the specific context of the situation. Hence such values baked into the technology design may not be sufficient to deal with questions that arise during deployment.

[37] drew attention to the need for designers to remain involved in the deployment of their technology, famously highlighted in his open letter titled *A Scientist Rebels* where he refused to share details of his technology design with *irresponsible militarists*. He goes to further illustrate how totalitarian governments or profit-seeking capitalists can ignore fundamental human values in their adoption and use of technology, the relevance of democracy in ensuring protection of citizen rights, and the responsibility of scientists to not be naïve and take responsibility for how their inventions and innovations could be used by others for unethical private or political gain. Similar views are expressed by [38] discussing the uncertainty with many new technological innovations in how they would influence future generations of humans, and hence he too emphasizes that usage principles should evolve continuously through oversight and monitoring during deployment. Our own views are shaped by the same logic.

[39] suggests that as part of the information age, a shift towards more responsible science and technology design and deployment will indeed happen; he says that the increased observability brought about in today's information age will make the agents act more responsibly. He then goes on to suggest what kind of ethical systems should guide the agents, and argues that individualist or *egopoietic* systems such as virtue ethics may not be sufficient to govern technology that affects societies, and therefore *sociopoietic* systems may offer a better foundation. Further, these would need to be linked to the environment through *ecopoietic* systems that take the needs of future generations into account.

However, whether the increased observability will lead to more responsible agents or not is questionable. This is illustrated by [40] in his articulation of the transparency-power nexus. Methods to bring about transparency to enforce responsible behaviour, may also be used by the powerful to control and discipline people. Observational control, ie. the ability to observe, can be used to regularize behaviour. Therefore, whether or not ICTs will lead to responsible outcomes may depend on the ethical principles with which power holders operate, which may in turn depend on the political systems in which they operate. We consider this view in the next section where we emphasize on power-based equality as a key ethical principle that should be followed by the designers and managers of ICT projects.

*2) Ethics in deployment management*

Given that ethical principles are clearly important in deployment management, just as much in design, this brings us to the question of how deployment feedback can lead to design changes. The action research and participatory design methods probably come closest to this.

Participatory design approaches are grounded in democratic values to enable users to influence the design, thus dealing with the challenge described by [34] of the gap between designers and users. However it does not formally state the need for an explicitly declared ethical grounding, and it still retains the objective as getting the design right as opposed to constant attention also being required for deployment management. Its relevance is also limited in today's context of how large ICT platforms are actually developed and scaled. Platform designers typically do not use participatory design methods at the outset because their incentives are driven by a build-and-break approach, with a goal to gain quick user traction through which they can claim access to more funding to scale their projects. Any fundamental problems in the design, irrespective of whether it arose because a participatory process was not followed or a consistent ethical perspective was not incorporated, therefore become concretized and hard to change as the platform grows larger. Efforts made later towards participatory design by consulting users are therefore not very effective and need to operate within the constraints of the original design. This inability to depart from a legacy design is a key problem for Aadhaar, for example. Solutions like virtual-ids which from a privacy point of view make it harder to join datasets amassed by different service providers [41], or offline authentication which has been proposed as a workaround to network failures for authentication at the point-of-service [42], can only alleviate some pain-points of centralization. They however still cannot solve the power imbalance created by a decision-making design that is centralized and hard for many people to appeal against.

Action research has more ambitious goals to continually shape the intervention based on deployment feedback, with all decision making done through the participation of the community in the process [43]. The framework we have stated can potentially be considered as a more practical approach than action research in our current context of widespread participation in platform based ICT projects, where users may not be given extensive privileges to shape the platform usage processes. Instead, we emphasize on the responsibility of the platform designers and managers to certainly take as much feedback as possible from the users, but also importantly acknowledge the influence that their own ethical systems have on how they choose to define the platform objectives and operational processes. Just like technology artifacts can create social order, technology platforms and their usage processes encode norms for user behaviour, hence it becomes important for the platform managers to recognize emergent usage norms and react in responsible ways to guide the formation and transformation of these norms by developing new features or processes, guided by clearly stated ethical principles. We further describe later that features and processes which empower the users to regulate usage norms on the platform, can be more desirable than a centralized management of usage norms. [44] describes this need for a dynamic application of ethical principles in a research project, than formulating an initial static ethics protocol that can be reviewed one-time by ethics boards. Our own thinking is on similar lines, albeit described for ICT implementation projects and not only for research projects undergoing an ethics review.

### 3) *Ethics and Artificial Intelligence*

Although we have discussed so far about ICT projects in general, the arguments we have made are just as applicable to ICTs that involve algorithms and data. Algorithms only codify the objectives defined by the designers, but the objectives themselves are an output of the underlying ethical system [45]. As an example, motivated by our responsibility for MV to ensure a diversity of viewpoints in the conversations happening on the platform, even if at the cost of user satisfaction, we have built content recommendation algorithms that enable platform managers to specify policies that can ensure (short-term) diversity and (long-term) fairness in recommending content [46]. The policies themselves can be defined by the platform managers based on what they value. In a similar way, machine learning classifiers have been shown to have biases emerging either from incompleteness in the data itself, or due to relationships between predictive performance and protected variables [47]. This leads to predictions or decisions that can be discriminatory. Methods have been developed to prevent this discrimination by articulating different kinds of fairness policies, which again emerge from varying ethical systems such as whether to achieve individual or group based fairness, or whether to only prevent discrimination or to also ensure affirmative outcomes for worst-off classes [45, 48]. Furthermore, these policies are also not important just at the design stage, but even while managing deployments there is need to ensure that biases in the data completeness are consciously addressed, and that models are attentively re-trained to continue to perform in line with the underlying ethical systems.

Honouring user rights pertaining to the data, such as privacy, anonymity, informed consent, and ownership, also need to be handled in similar ways, by reasoning about the underlying ethical systems of the technology designers and managers [49]. Similarly, accepting accountability for the outcome of the algorithms, transparency and explainability of the results, and providing appeal procedures against decisions made by the algorithms [50], are necessary to deal with mistakes and take corrective action. Participatory methods can help develop strategies that are in agreement with the users [51], but as argued earlier to preserve continued ethical functioning of the projects will also require due attention during deployment, such as to ensure responsible functioning of appeal processes, or getting continuous user feedback to detect problems and evolve mechanisms to shape usage norms on ICT platforms. Ensuring ethics by design in artificial intelligence technologies, as adopted in declarations such as ICDPCC [52], are therefore unlikely to be sufficient by itself. Broader frameworks such as the one we have proposed will be needed even for projects that have a strong component of algorithms and data, to pay attention to both design and deployment, guided by the underlying ethical principles and politics of the designers.

## III. Power-based Equality

Given the importance of ethical underpinnings in the conceptualization framework described above for ICT projects, we next discuss the relevance of a key value that any project should embrace, of working towards power-based equality among the actors directly or indirectly affected by the project. Our claim is that this can help distinguish between projects that empower people and those that disempower. Projects that reinforce existing power relationships or create new power relationships towards a few, or towards the technology itself, end up being disempowering.

### A. *Relevance of a power-based equality principle*

Disempowerment seems to result from several recurring negative patterns, and which are seen more widely than in just ICT projects. One such pattern is the mindset of governments that legibility and simplification as a means of control and coordination of the population is unequivocally useful [53]. This often leads to the design of regimented programmes that suppress the ability of people to flexibly solve their own problems, which disempowers them, and eventually even fails to produce meaningful outcomes. This pattern has been seen in many initiatives. The standardized blueprint imposed by the Indian government to operate Internet-based information services kiosks in rural areas, became a constraint for kiosk entrepreneurs to be resourceful in diversifying their services and finding workarounds for technical glitches that they encountered when operating according to the standard procedures [12]. This even led to several kiosks becoming unsustainable. Similar rigidity imposed by the technology driven design of the Aadhaar system in India has actually made it harder for welfare dependent low-income citizens to engage with the state, because Aadhaar's centralized processes essentially eliminate the civil society from intervening who could have provided assistance to the people in interfacing with the technology driven systems, thereby actually disempowering the poor and their community institutions [31, 32].

Another pattern is the strong belief that competitiveness among people, operating within a laissez-faire framework of minimal regulation and external coordination, will create conditions for equal opportunities of growth for everybody [54]. This however in a world that is *a-priori* unequal leads to unfair conditions for competition [55], and reduces the value placed upon cooperation and regulation to create fairer conditions for equitable growth in an unequal world [56]. It also ignores the role that regulation can play in addressing the root causes behind inequality such as the unequal distribution of skills and opportunities, which unless addressed directly to build skills and mentor people will only further increase inequality and disempowerment. This pattern occurs frequently in many contexts. The lack of attention paid in most social media platforms to the regulation of user behavior and creation of norms for responsible usage has led to incidences of fake news that have even subverted democratic institutions [20]. Recommendation algorithms that can help regulate user behaviour to some extent, are rather built to drive user engagement in most profit making social media platforms towards sensational and alarming information [58], rather than being based on cues about the authoritativeness and completeness of information that can meet social objectives such as user learning. This has broken the myth that social media can disrupt powerful gatekeepers and democratize the ability for anybody to make themselves heard; rather social media has given rise to new forms of agenda setting and mechanisms to drown even legitimate voices [57]. In a different context, the emphasis by microfinance institutions for the poor to just focus on financial metrics has meant that hardly any attention is paid to mentoring the borrowers to effectively utilize their loans [59]. This not only reduces the effectiveness of such programmes, but also allows more skilled people to get further ahead. The attraction to individualized concepts like universal basic income and cash transfers instead of support for collective efforts, can also be attributed to this pattern with its focus on individuals, competition, and absence of regulation.

A third pattern is the nature of capital to centralize itself, exploit existing inequalities to its advantage and thereby reproduce them [55], and of capitalists to further use their power of capital to subvert any regulatory efforts made by the government or media to impose fairness constraints [56]. This pattern reinforces the first two patterns through a tight nexus: Legibility enhancing programmes of the government provide tools to capitalists to increase formalization and create new spaces for capital transactions; this increases the opportunity to create wealth; due to unfair competition the wealth gets further concentrated; this concentrated wealth is able to influence the public and the government to draw attention away from the need for regulations to create an equitable distribution of opportunities; as a result the status quo is retained with an emphasis on individuation and competition instead of collectiveness and cooperation, allowing for continued exploitation and perpetuation of inequalities.

These three patterns ultimately create undesirable power structures. We define power as the ability for an actor to continue to successfully influence their environment according to their will [61]. Thus, for technologies or standard operating procedures created by the government, their power lies in ensuring that other actors operate in accordance with the protocols laid down for the system. Algorithms embedded in the functioning of social media platforms have power in influencing and controlling the behaviour of their users according to the objectives defined in the algorithms. Social media users also have power depending upon their skills and connections to influence which information gets shared or blocked according to the strategy they want to impose. The logic of capitalism lies in enabling power holders to exercise their power to ensure their own survival, be it through paying lower wages to employees, or influencing policy for less regulation, or influencing media to create policy legitimacy or to shape consumer preferences. Further, the power of an actor is not only their ability to influence the environment, but to continue to influence it, ie. exercising power does not lead to it being challenged or reduced. Getting away with not equitably creating skills and opening up opportunities for growth, and not encouraging structures for cooperation and collectivism, helps ensure that power remains consolidated in existing structures and is not effectively challenged or dissipated.

The three patterns we have described are common ways in which undesirable power structures are created and entrenched, and due to these power-based differentials between different actors they lead to undesirable outcomes or a reduced effectiveness of the programmes and development initiatives. Incorporating a power-based equality principle as a core ethical value for the design and management of ICT projects, may help ensure that ICT projects do not incorporate these patterns themselves, and potentially even counter the occurrence of these patterns in other systems in the world. ICT projects indeed are embedded in an ecosystem of actors who manage the ICT systems and processes or are affected by them. These actors have pre-existing relationships with one another of power or cooperation or other dynamics. With power-based equality as an underlying principle, ICT projects can very well aim to alter these dynamics. Projects like Mobile Vaani have given power to the people to protest against the poor delivery of public services, and put media pressure on the authorities to act upon the grievances [22]. Social media platforms have given power to the people to coordinate the formation of collectives and joint action [62]. The availability of information such as market prices of agricultural commodities for farmers has reduced information asymmetries by giving more bargaining power to the farmers to get better prices from the traders for their produce [63, 64]. We therefore need a method to identify whether or not an ICT project is based on power-based equality embedded as a principle in its underlying ethical framework.

Assuming that based on the examples we have given above it is indeed problems to do with the distribution of power that lie at the heart of the issues we have discussed so far, the question it raises is: Can we build a modelling method for the design and management of ICT projects to assess whether a project will bring about (or is bringing) power-based equality among the actors? Such a modelling method could help distinguish between projects that empower people and those that disempower. We come to this modelling method next, to incorporate power-based equality as a principle in the underlying ethical framework for ICT projects. Referring to Figure 1, this modelling approach can be applied at the third part of the design layer, to specify the different actors who would directly or indirectly participate in a project, their mutual relationships, the distribution of power among them, and how the ICT project could change the power distribution. It can also be applied in a concurrent manner to guide deployment management, by keeping track of the power distribution dynamics unfolded by the project.

### B. *Modeling methodology*

We suggest a modelling approach that is inspired by the cybernetics [65] and systems-thinking methodology [66] which examines the system as a whole made up of many parts that interact with one another based on various rules and lead to certain systemic behaviours. The model should allow designers and implementers to specify the various actors and their relationships in a structured manner to make these legible. Once the system is expressed in a legible form it becomes amenable to analysis, reflection, corrections, and more detailing to bring it closer and closer to the real world [61]. We only go as far as suggesting a framework to express system dynamics, than to predict behaviour; behaviour prediction models can potentially be built on top of the framework. However, we outline certain system archetypes which often lead to undesirable outcomes, and can be spotted in the models.

We propose modelling an ICT project in terms of its *actors*, *resources* possessed by the actors, *activities* performed among the actors, and *decision functions* governing the activities. Actors may be people, organizations, collectives, or even technology artefacts and processes, involved in a project. These actors may possess resources such as information required to make decisions, know-how required to execute activities, discretionary or veto rights to make decisions, etc. Activities may include services performed by an actor for another actor, and which may consume or produce resources. The execution of activities can be controlled by decision functions based on the resources possessed by actors involved in the activities. An example for Aadhaar based service access is shown in Figure 2a. There are three actors in this system: A user, the Aadhaar system which accepts/denies user authentication, and the actual service availed by the user. Users possess resources such as know-how and capability to engage with the Aadhaar system to operate it successfully. Those with less know-how may face problems such as with rectification of Aadhaar registration errors or to deal with situations when technology failure may cause authentication errors leading to service denial. The activity of Aadhaar based authentication is therefore governed by a decision function that is dependent upon the know-how of the user. The activity for availing the service is governed by the output of the Aadhaar authentication, and of course whether or not the user is entitled to the service.

This modelling in terms of actors, activities, resources, and decision functions, is able to capture the three kinds of patterns discussed earlier. Regimented programmes with less discretion for bypassing protocols would look like the Aadhaar example shown in Figure 2a, with star-shaped network structures and sequential activities that would indicate fragile networks with a single point of failure resting with the process as an actor. In contrast, Figure 2b shows more meshed and connected networks where community institutions can support people to access services. ICT platforms with minimal regulation and oversight would look like the social media platform shown in Figure 3a, where there are no decision functions imposed on users to communicate with one another. Figure 3b on the other hand shows a platform where users themselves impose regulations on one another for responsible usage of the platform. Programmes like income support to individuals would bring changes as shown in Figure 4a, where initial inequalities in the resource distribution of skills are enhanced by the programme. In contrast, Figure 4b shows that programmes which support skills building will not only lead to more effective socio-economic development but also result in more equitable outcomes. Online grievance redressal mechanisms as followed in many government schemes, look like Figure 5a where service providers can impose significant discretion on redressing grievances. Regulatory loops imposed by media systems as shown in Figure 5b, can keep this discretion in check.

These models can be analysed to identify three system archetypes that can lead to different kinds of power effects. Connectivity metrics about the network structure to assess resilience to edge or node failures, can indicate whether the model has only a small number of important points; these are likely to be the loci of power with centralized decision making privileges. Examining the distribution of resources across the nodes can indicate whether the resources are equitably distributed or not; inequitable distribution is likely to lead to further inequity. The presence of regulatory loops can be spotted by identifying cycles in the network to check whether decision making links emerging from powerful nodes are countered by other nodes; this can impose checks and balances on power concentrations. Thus, we can identify at least three kinds of archetypes: concentration of power among a few actors, inequity in the distribution of resources (contributing to power) among the actors, and the absence of regulatory loops to keep a check on power. The existence of these archetypes can be spotted with the modelling approach we have outlined here for ICT projects, to evaluate whether or not undesirable power effects could arise or are emerging through the projects.

Our key goal behind coming up with a modelling approach is so that the models can be compared with one another, at the design stage to choose one model over another that appears to favour more power-based equality, and at the deployment stage to monitor the evolving power dynamics in the system. A comparative analysis of the models can be qualitative to begin with, but with more detailed modelling even quantitative metrics can be developed to compare the relative merits and demerits of different models. Further, the models may not just be analysed statically in terms of their configuration, they can also be turned into dynamic models governed by state transition equations which can be simulated to observe the effects over time. As an example, in Figure 6 we outline a dynamic model about how a participatory media platform deployed in a community builds its credibility. Articulating the model raised new questions about what the credibility function should look like, or how much minimum credibility should be attainted for the media to become effective in exercising its influence; this is precisely the role served by models to identify places of over-simplification that need more detailing, which eventually leads towards a better understanding of the system.

The systems-thinking approach has conceptualized rich insightful methods that can be used here to measure and project the system dynamics, and can be helpful for managers to make decisions and to even provide reasons behind their decisions for review by other stakeholders [67]. This approach can therefore help answer questions of whether to choose a particular model or not, whether the model will find an equilibrium when unrolled over time, whether the regulatory loops are strong enough, etc. The models can be made as simple or complex as needed, to answer the questions that are put up to them.

We have shown so far that several negative patterns that frequently lead to undesirable outcomes for ICT projects, can actually be explained through a common framework of power effects, and that these effects can be anticipated or tracked through the modelling approach we have proposed here. The modelling approach can help examine ICT projects to evaluate both at the design and the deployment stages whether power-based equality forms a key principle of the underlying ethical framework for the projects. This can be done by looking out for at least three system archetypes that we have identified: Whether power is getting concentrated in a few hands, whether regulatory loops are in place to keep a check on such power concentration, and whether the underlying resources that contribute to power are equitably distributed or not. We next briefly discuss the special case of power given to technology artifacts directly, given the rise in algorithmic decision making and an oft-observed belief in the correctness of technology. We then discuss our proposed power-based analysis methodology, in relation to other frameworks proposed in the literature.

### C. *Power to technology*

We have discussed examples earlier such as Aadhaar and Facebook's centralized technology driven architecture for decision making – authentication in the case of Aadhaar, and decision making about the acceptance of permissible speech on Facebook. The technology artifacts, or processes driven by the technology artifacts, emerge as key actors possessing concentrated power according to the modelling method we have proposed. To keep a check on this power, other actors too need to have power to appeal against the decisions, and have access to explanations about the decisions. These requirements have been noted in recent declarations for ethics by design in the use of artificial intelligence technologies [52], but as argued earlier, this needs to be incorporated at both the design and deployment stages. The modelling method can be used to examine which actors have access to resources required to keep a check on power assigned to technology and process artifacts.

It is worth discussing other non-platform forms of ICTs as well, such as IOT technologies projected to improve agricultural productivity [68], or big-data based approaches such as through the use of satellite data and other large datasets to make farming recommendations [69]. Reliance on these technology artifacts is putting more power in the hands of the artifacts themselves, and we argue that such arrangements even in non-platform ICTs need to be handled in a better manner for three reasons. First, several limitations have been noticed about these technologies, such as not having enough data about the local context to fine-tune the recommendations, or the lack of transferability of models across different contexts, which can lead to mistakes [70]. Methods to support more equitable distribution of power would suggest the need for similar mechanisms as discussed earlier for Aadhaar, like the explainability of the recommendations so that users can decide whether to trust them or not, and to encourage possibly the users themselves to provide more local context that can help improve the algorithms. In other words, to not let technology dictate decisions but only provide reliable supporting data to the users to make their own decisions.

The second reason is related to power relationships that get established between the owners of the technology and its users. Whether the owners of these ICTs can misuse their power, such as by providing privileged access to data to other actors like traders or insurance providers, can reinforce the power imbalance between the traders and farmers, or insurance providers and farmers. Direct comparisons can in fact be drawn with ICT platforms such as Facebook where very similar concerns have arisen. In fact, for technologies that rely on data provided by the users to improve the technology, if the users can be compensated in some manner it can neutralize power differentials between users and owners of the technology.

The third reason is about power relationships in the ecosystem that are altered by the ICTs: Whether technologies like IOTs can be afforded by everybody or only by large farmers, will determine if the ICTs can help remove existing inequities or not [71].

We therefore argue that wider ecosystem modelling is needed to understand the power dynamics affected by giving power to technology itself. Power should be given to technology only when its introduction helps reduce power-based inequity in the wider ecosystem, including between the technology and its users, the technology owners and the users, and between the users themselves. With this view, open-source systems running on distributed infrastructure with appropriate data management tools for privacy, deployment guidelines for capacity building of users to ensure equitable access, and with objectives to provide information or services to counter existing power inequities in the ecosystem, may seem to be more reliable guidelines to design ICT systems that can avoid undesirable outcomes.

### D. *Related work*

Our modelling approach in terms of actors and links between actors, is very similar to ANT (Actor Network Theory) [72]. ANT helps explain why some networks are stable but others do not sustain, by understanding the aligned interests among the actors. ANT however does not allow a modelling of overall system objectives, and does not define any specific patterns that could lead to power differentials among the actors. The systems-thinking based modelling approach proposed by us, and the list of archetypes that can lead to undesirable power effects, can be used to describe these aspects missing in ANT.

The benefits in the ICT4D space of the systems-thinking approach of seeing a system as a whole, is discussed by [73]. It helps to see technology in a wider context of social systems comprised of different kinds of actors who interact with one another. Useful concepts such as open and closed systems, and boundaries of a system, can help determine the extent of complexity that was chosen to be modelled, and consequently remain aware of what was not modelled that could lead to surprises. Concepts like functions that relate inputs with outputs, and composition of functions in dynamic systems that could lead to emergent effects, helps bring precise thinking to the function definitions and assumptions therein, which can be a useful exercise to bring forethought in deciding actions. Positive and negative feedback loops are another useful construct to keep a check on emergent phenomenon. Decomposition of large complex systems into smaller hierarchically organized independent sub-systems, is also a useful technique to simplify the models. Our proposed modelling approach can readily benefit from such techniques developed in the systems-thinking literature.

The capabilities approach to studying the effect of ICTs is another modelling approach to which we can draw similarities [74]. It identifies the need for people to possess essential capabilities that can help them make use of opportunities, and equality in capabilities therefore emerges as a key concept. This is similar to our own insight about the need for equitable distribution of resources. The capabilities approach however does not suggest any modelling methods, especially something that can be used to analyse the presence and effects of regulatory loops as well.

In a similar way, power as a concept has been studied extensively, but not with a rigorous modelling approach such as what we propose. The social sector has utilized power analysis since many years as a tool to help communities understand different kinds of power dynamics around them [75, 76]. They distinguish between different expressions of power, as *power-over*, *power-to*, *power-with*, and *power-within*: An actor may have *power-over* other actors to bring about certain outcomes, actors may have the *power-to* do their will, actors can build *power-with* one another through collectives, and actors can have *power-within* themselves based on their individual or collective self-efficacy. These expressions of power can be made in different forms that might be *visible*, *hidden*, or *invisible*. *Visible forms* are like written down formal rules and procedures that may reveal expressions of power, *hidden forms* are when power is expressed by exercising influence and setting agendas that are unwritten, and *invisible forms* are when dominant ideologies or norms may govern the expressions of power. Various types of expressions of power and their respective forms may be made in spaces that are *closed*, *invited*, or *claimed*: *closed spaces* are where decisions are made behind closed doors in a non-transparent manner, *invited spaces* are where people are especially invited to participate, and *claimed spaces* are when less powerful people come together to create their own space. Each of these spaces may impose different checks or make allowances to the expression of power. The spaces themselves may operate at various levels such as at the *global*, or *national*, or *community*, or *family*, or *individual* level. This taxonomy has been found to be useful to help communities discuss and write down how they see power being exercised in their lives, and how they may intervene to alter the power dynamics for social change. PowerCube [77, 78] and NetMap [79, 80] are popular tools that are used for such community inquiry processes. PowerCube is useful to list down the different power relationships, while NetMap takes a social-network based approach to identify different actors, relationships between the actors, and the influence each actor may hold. Although these methods are useful to identify and categorize different kinds of power relationships, but they do not go as far as developing a rigorous model that can be used to identify undesirable archetypes or specify dynamic relationships.

It is worth noting that much of this taxonomy of power described above, and are own formulation in terms of resources and decision functions to govern activities, are just operationalizing methods for various concepts of power that have been discussed and debated for a long time. Marx and Engels' concept of false consciousness [81], and Gramsci's notion of hegemony [82], are examples of invisible power exercised through propaganda and creating dominant ideologies that impede people from realizing the underlying mechanisms behind their exploitation. Foucault emphasized on knowledge as a means to challenge the legitimization of power, and to overcome the disciplining mechanisms when power is wielded specifically to suppress the realization of people that they are being controlled and manipulated [83]. Scott discusses how resistance to power is also seen through small events of non-cooperation by ordinary citizens, and how these may transform into larger forms of protest [84]. Relationships can be drawn between these theories and our operationalizing methods, for example, the concept of hegemony is related to invisible power, which can be challenged by building knowledge as a resource, and incorporating regulatory loops to keep propaganda in check. Overall, our modelling approach in terms of who has power to influence their environment, resources that contribute to this power, and regulatory loops to keep power holders in check, seem to be consistent with literature that has discussed different concepts of power.

An opposing view about the insufficiency of modelling is discussed by [85] who argue that systems-thinking and cybernetic based approaches, along with other *managerial* approaches, are not sufficient to model complex social systems. They claim this in the context of a broader argument that societal problems are wicked problems to solve through a planning approach, such problems cannot be solved in entirety but only continually re-solved. Such views should be kept in mind when using systems-thinking based approaches, to identify their limitations and improve them potentially through more complex or context-specific models.

## IV. Socialization Of Project Teams

We have shown so far that an ethical framework needs to provide an underlying foundation to define the objectives of an ICT project, its design, the management of its deployment, and ensure that feedback from the deployment is conveyed to fine-tune the design. Further, power-based equality as a key ethical principle can be important to ensure that responsible outcomes arise from the project. These ethical principles are put into action by the project teams, when they are designing or re-designing the project, or shaping its usage norms through careful management of the deployment. Given the importance of human agents in the process, in this section we describe various aspects that may impact how well people in the project teams implement the ethical principles in their day to day work. Most of our arguments are based on our own experience with working with Gram Vaani for over a decade, which built the Mobile Vaani platform.

We outline at least four aspects that seem to be relevant. First is the organizational or team structure, in terms of whether it enables the sharing of insights between different team members. This is clearly easier in small teams. As teams grow the common way to organize them is along different functions. This can however become restrictive in information sharing across formal functional boundaries that get created as a result of the segregation between teams. We experienced this closely at Gram Vaani. As MV grew, we built function-specific teams for content creation, moderation, field team management, engineering, etc. While this helped the teams build functional specializations, it slowed innovation and quick reactions to feedback shared between various teams. For example, with a smaller team that spent a lot of time with one another, any observations made by the moderators about the quality of voice reports recorded by different volunteers would reach the field team quickly. The field team was then able to guide volunteers in customized ways to record better content. While this feedback sharing happened organically and informally in small teams, it took us a while to realize that as the teams grew and functional segregation increased, this feedback sharing reduced. A specific process ultimately had to be formalized for this purpose, with the institutionalization of regular calls and meetings between the teams to exchange insights. Identification of many such informal processes, followed by the formalization of these processes, and then a challenging transition phase to move from informal to formal processes, has been an ongoing activity at Gram Vaani as the organization has grown.

Functional segregation however has other more serious effects than just to impede the flow of useful information. For example, we found field teams to be empathetic to problems faced by users and volunteers in using the platform, such as technical issues like call disconnections while recording content, or even the need for capacity building to effectively make use of the technology. Perhaps this empathy emerged because the field teams directly faced the users and volunteers, and felt responsible to guide them in the use of the platform. However the technology team, and increasingly the moderation team, who were hardly directly in touch with the users, seemed to lose this empathy as the functional segregation increased. This was noticed in terms of slower evolution over the years of user-facing help features in the technology, and less frequent guidance calls given by the moderators to the users and volunteers for content recording. Arguably some of this also happened because of competing priorities to build other features, or organizational resource constraints that restricted continued investment in user capacity building, but the fact that these issues rarely got discussed across the organization is probably because the issues did not make their way out from the silos in which the field team was operating. A similar issue seems to have happened at Facebook, as stated in various news reports, that signs of data breach and platform misuse were not heard by different teams and handled in priority [26]. Socialization between teams therefore seems to be essential not only to share feedback, but to also share values that are important to different teams, or in other words to bring a consistency in the ethical system within which different teams operate. In places like Facebook or Gram Vaani, the diversity across teams coming from different academic backgrounds and professional experience is actually an asset, of being able to look at problems from different perspectives, but mutual interaction and discussion is essential to utilize this asset.

The second related aspect to having teams respond based on a common ethical system is the organizational mandate itself. A clarity in this mandate, along with socialization and sharing of feedback between diverse teams, can potentially impose consistency and rigour in following an ethical system for the organization to design and run an ICT project. At Gram Vaani, eventually this realization of the need to support capacity building of users to utilize the platforms effectively, did lead to changes in the operations and priorities of various teams. Evidence however points to cases where deliberate ambiguity is created among team members in having a common organization-wide view [86]. Internal propaganda seems to be used to create an impression, for example, for engineering teams about world-changing impact that their work is having, and isolating them from business teams who have a closer ear to the ground about potential violations of user rights that may be occurring to satisfy the business objectives of the organization. Workplace segregation and having different reporting chains for different teams seem to historically have been common strategies used to create segregation to prevent unionization and collective action, and which similarly is able to evade a reconciliation of differences in views between different teams. This allows ambiguities to persist, even though inconsistencies can lead to undesirable outcomes.

A third aspect that influences choice of the ethical system, at the team level or organizational level, is the political ideology of the team members. When this deviates from the ideology of the users, it again opens up faultlines to build and run ethically consistent ICT projects. We take the case of blue-collar gig economy platforms such as for drivers and couriers. ICT project teams on such platforms are largely comprised of a white-collar workforce of engineers, designers, project managers, business development executives, etc, who have had increasingly divergent views from blue-collar workers [87]. White-collar workers tend to be less opposed to inequality, more drawn towards personal grievances than collective grievances, and less inclined to participate in unions, than blue-collar workers [88, 89]. Initiatives like the Tech Workers Coalition [90] are trying to bridge the divide, but until such time differences in political ideology will directly influence what values and outcomes the project teams may prioritize. Reports like the user interface design of the Uber app for drivers to nudge them to keep driving [91], and setting inhumanly difficult incentive targets for drivers [92], are clearly outcomes of having altogether different political ideologies between the platform designers and managers, and the drivers. Further, while Uber employees earn high salaries, the drivers who are considered as private contractors have seen their earnings gets constantly squeezed, and no significant voices of Uber employees seem to have been heard so far about this issue.

The same gap in ideology may also arise with other ICT platforms where the project teams and project users are different from one another. This is evident in the case of Aadhaar, where there is a clear divergence in the views of the technologically minded architects of the platform, and views of many users represented by the civil society about problems with the platform [3, 31]. One side with a strong sense of high modernity seems to believe in the utilitarian principle of *greater good* with failure cases regarded as a minor statistical error, while the other side gives prominence to the seriousness of this statistical error which still represents several million people and has allegedly had grave effects such as even starvation deaths caused due to denial of welfare benefits arising from technology or process failure. Media propaganda and dominant business practices, often shaped by the wider political economy nationally and even globally, further influence the ideologies of the ICT project team members.

It is therefore worth spending some time to discuss the political economy of technology, which may help explain such divergent views. Most technologies require a large investment of capital for their development. This includes the setting up of telecom networks, storage platforms in the cloud, computation infrastructure, applications and algorithms to operate on the infrastructure, etc. Consequently, a significant need has to exist or be created for purchase of the technology. This is done in many ways. In the ICT4D context, we saw through an analysis of mass media in India that governments, corporate, and the media were aligned in projecting an optimistic and aspirational view about technology in bringing change [93]. This manufactures democratic consent for legitimization of even those technology policies that can actually be disempowering for many people. The state is thus able to build better tools to monitor and control the population, they are able to justify for political gain that they are actually bringing positive social change through technology, and corporations are able to find a customer for their businesses. In a wider context of economic policies in general, we saw through an analysis of mass media and parliamentary question hour data that any constituencies harmed by policy choices could make themselves heard only if their cause was politicized, and even then rational and informed responses in legislation were not common [94]. Rather the debates would often devolve into political blame games without a deep introspection and understanding of the details by the law makers. Thus a clear nexus or mutual understanding between technology companies, the state, and the media, about a case for greater use of capital intensive ICTs for development, along with suppression of views and politicization efforts by the civil society about undesirable outcomes arising from the technology policy choices, builds a technology optimistic outlook among ICT4D project design teams. This might differ from the views of the users and hide the complexities in realizing positive outcomes from ICTs [95].

The non-ICT4D context operates similarly. Marketing becomes an ally of the technology companies, both for creating a want for their technologies among the people, as well for other products that are advertised on digital ICT platforms and contribute valuable revenue to the platforms. Furthermore, digital platforms seem to have greater advertising efficacy due to their targeting capability based on precise user knowledge to achieve a higher return on investment [96], and therefore digital ICT platforms and marketing become mutually reinforcing of one another. This provides the crucial consumerist fuel for economic growth, especially for investors who are looking to invest their capital in new markets and opportunities. This capital which is said to be in over-supply [97], chases any opportunities that exhibit early success, irrespective of any ethical foundations for the conceptualization or operations of the technologies. Such a dominant view of action and change through technology also ends up permeating the ICT teams in technology companies, and seems to have been impacted only recently after alarming personal experiences of the teams as users themselves, or as friends and family of concerned users [98]. Hence the political ideology of the teams, which is shaped by the wider political economy of technology, manifests itself in the design of ICTs.

The fourth aspect is power relationships between the teams. Organizations are typically organized hierarchically, both within teams and also to enable communication across teams. Power biases created due to these hierarchies can lead to some views getting suppressed and ignored. Organizational policies are therefore needed to ensure that employee voice is heard and acted upon. It is interesting that at Facebook, according to news reports, even when an organizational mandate by the leaders was shortcoming to handle the problem of misinformation campaigns, it was actually a handful of employees who came together and set up a taskforce to address the problem [26]. Mechanisms like co-determination practiced in Germany [99] which give employees a board seat, can legitimize such bottom-up methods to ensure that ethical frameworks are clearly defined and implemented. Similar asks have been put forth to build user associations that can govern ICT platforms based on priorities defined by the users [100]. With ICT platforms being used by millions and billions of people, yet designed and managed by only a handful of people, the need for such representation is perhaps justified both for accountability as well as for democratic reasons.

In summary, we argue that ultimately having an ethical framework for the governance of ICT projects depends on the project team members, and their ability and inclination to do it is shaped by aspects such as the organizational structure for inter-team interactions, clarity in the organizational values, political ideologies of the project teams, and power relationships within the team. This shows that ICT projects which otherwise appear to be entirely technologically driven, and increasingly so with AI based automation, are actually influenced a lot by the organizational culture and its values. Organizations with a strong culture of communication and respect for their team members and for users, are likely to design and manage ICT projects more responsibly as compared to organizations that may not have such practices already in place. As ICTs become more and more pervasive, and bring the promise of scalability and intelligence, the fact that ultimately responsible outcomes depend on the organizational culture is a humbling reminder of the importance of values with which organizations are built and run. If an organization is not foundationally strong on these aspects, it is unlikely that ICTs can fix those weaknesses, rather the weaknesses could manifest themselves even more strongly if the ICTs reinforce existing power relationships or the limitations of the ICTs are not well understood, leading to undesirable outcomes.

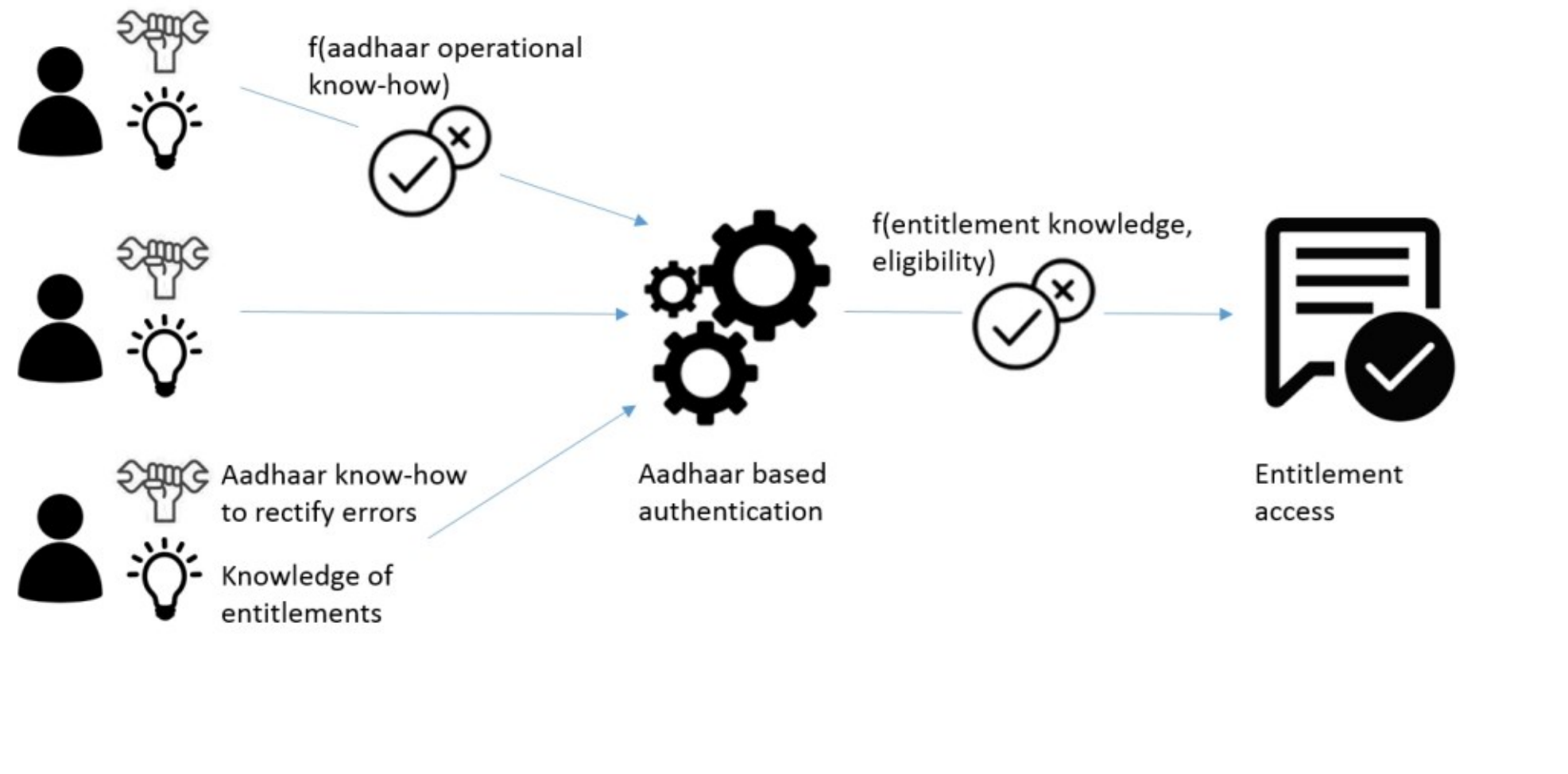

Figure 2a. To the left are shown people seeking access to entitlements. An Aadhaar-based system is deployed for authentication. Only successfully authenticated users can get their entitlements. People need resources of knowledge about entitlements, and know-how to rectify errors such as spelling mistakes in names or addresses in their Aadhaar enrollment, to avail entitlements. Decision functions shown as f(..) reveal whether know-how and knowledge, and eligibility of people, will lead to access.

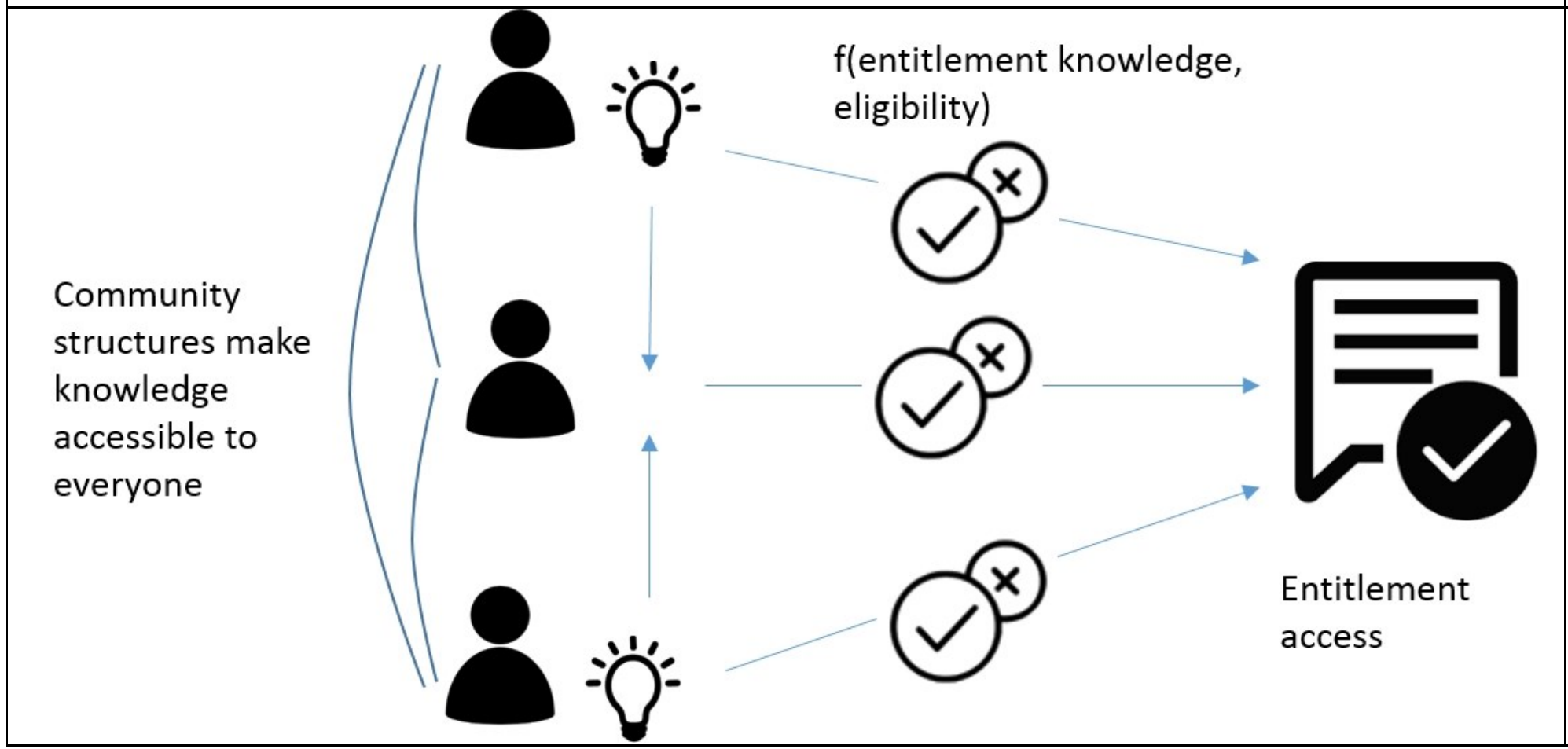

Figure 2b. In Figure 2a, if somebody does not have the requisite know-how to engage with the Aadhaar processes, they can get locked out from access to their entitlements. In Figure 2b we show a scenario without Aadhaar, where community members are able to help one another acquire the knowledge or follow-through with the processes, to help the community members get access to entitlements. Although not shown here, this also helps build institutional capability locally in the community.

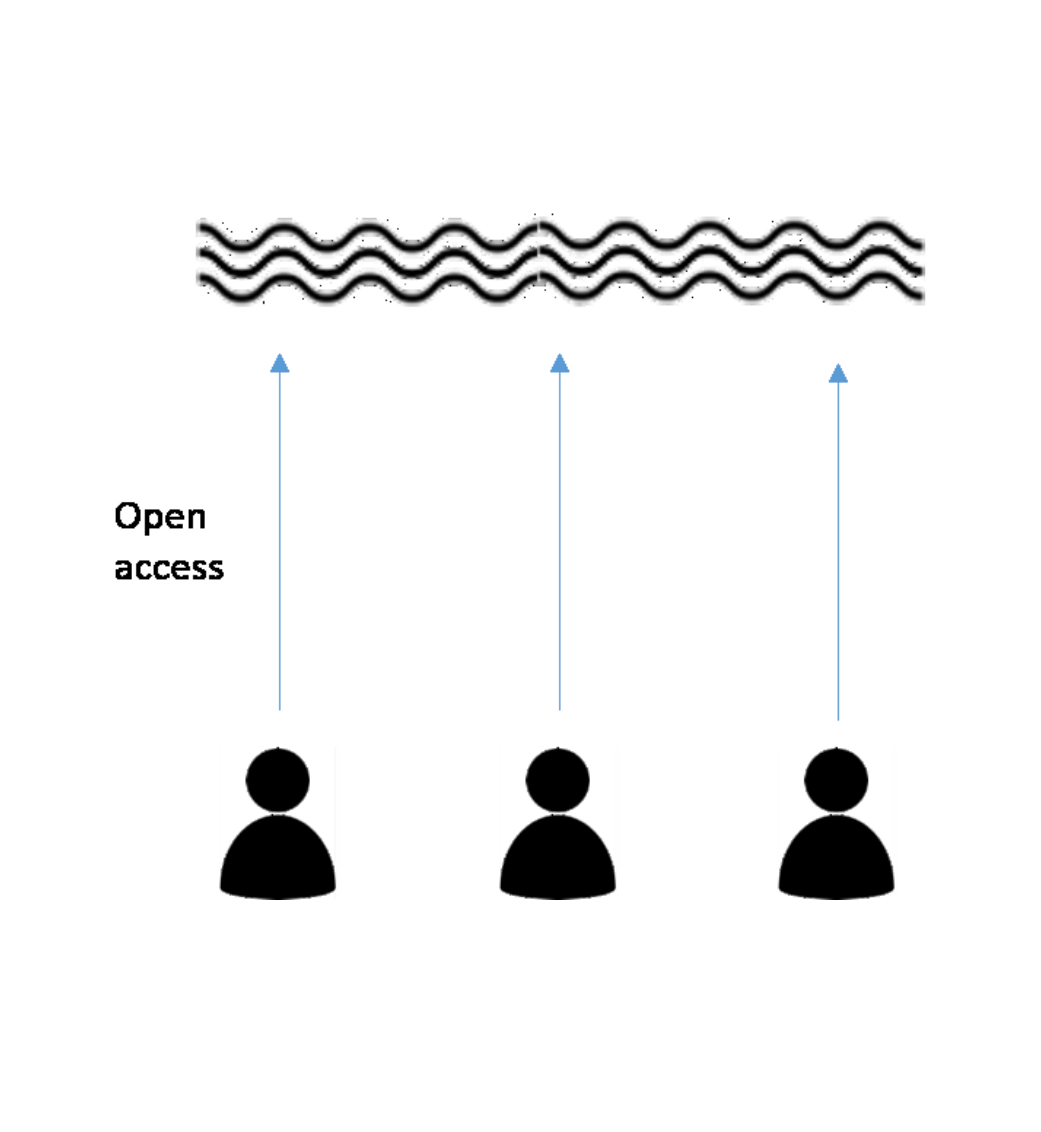

Figure 3a. This shows a simple setup where anybody can access a given communication medium to interact in the group.

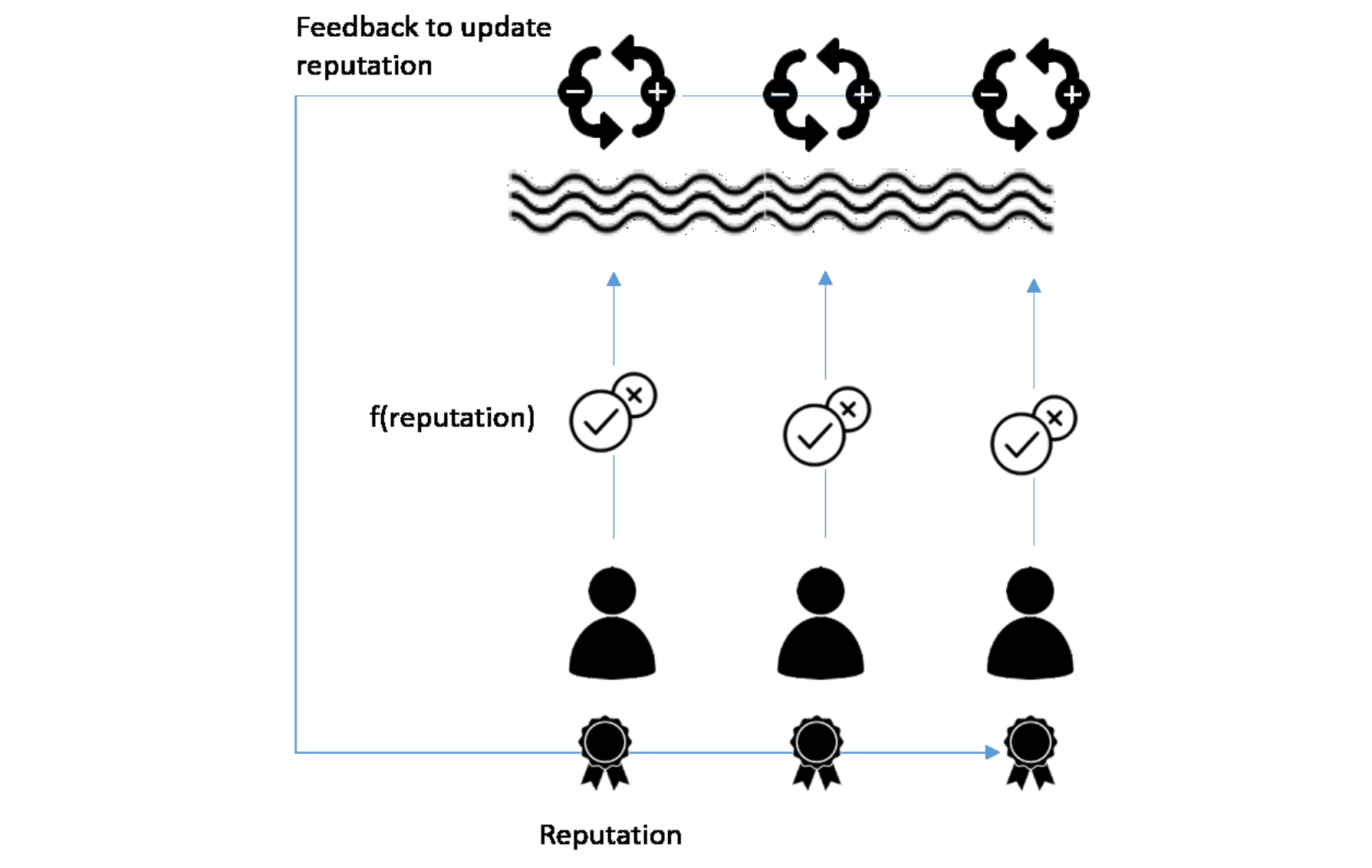

Figure 3b. This shows a setup where access to the communication medium is controlled by the reputation resource of users. The reputation itself is calculated through feedback by other users about prior submissions made by a user.

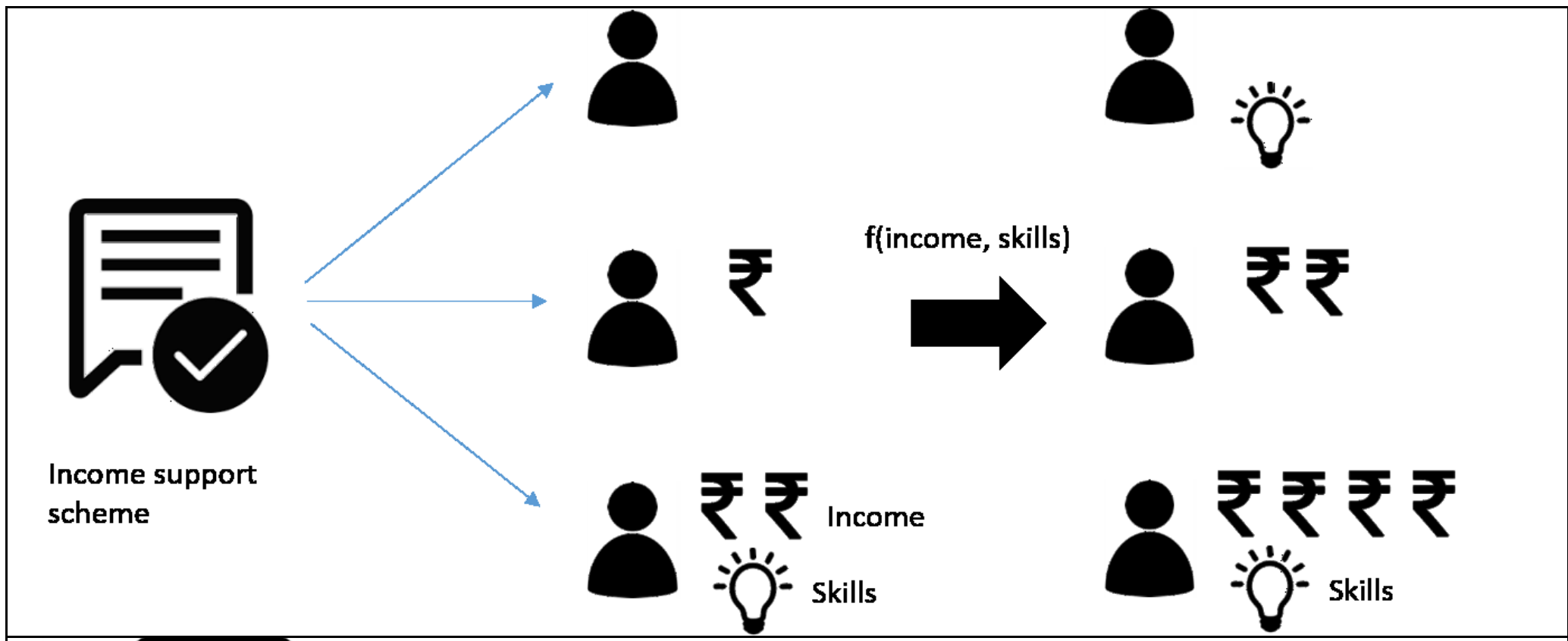

Figure 4a. Shown here are the effects of an income support scheme on people who have resources of different skill levels and incomes to start with. With access to additional income from the scheme, people who were better off at the start are able to do even better, while others who did not have the required skills see varying outcome effects.

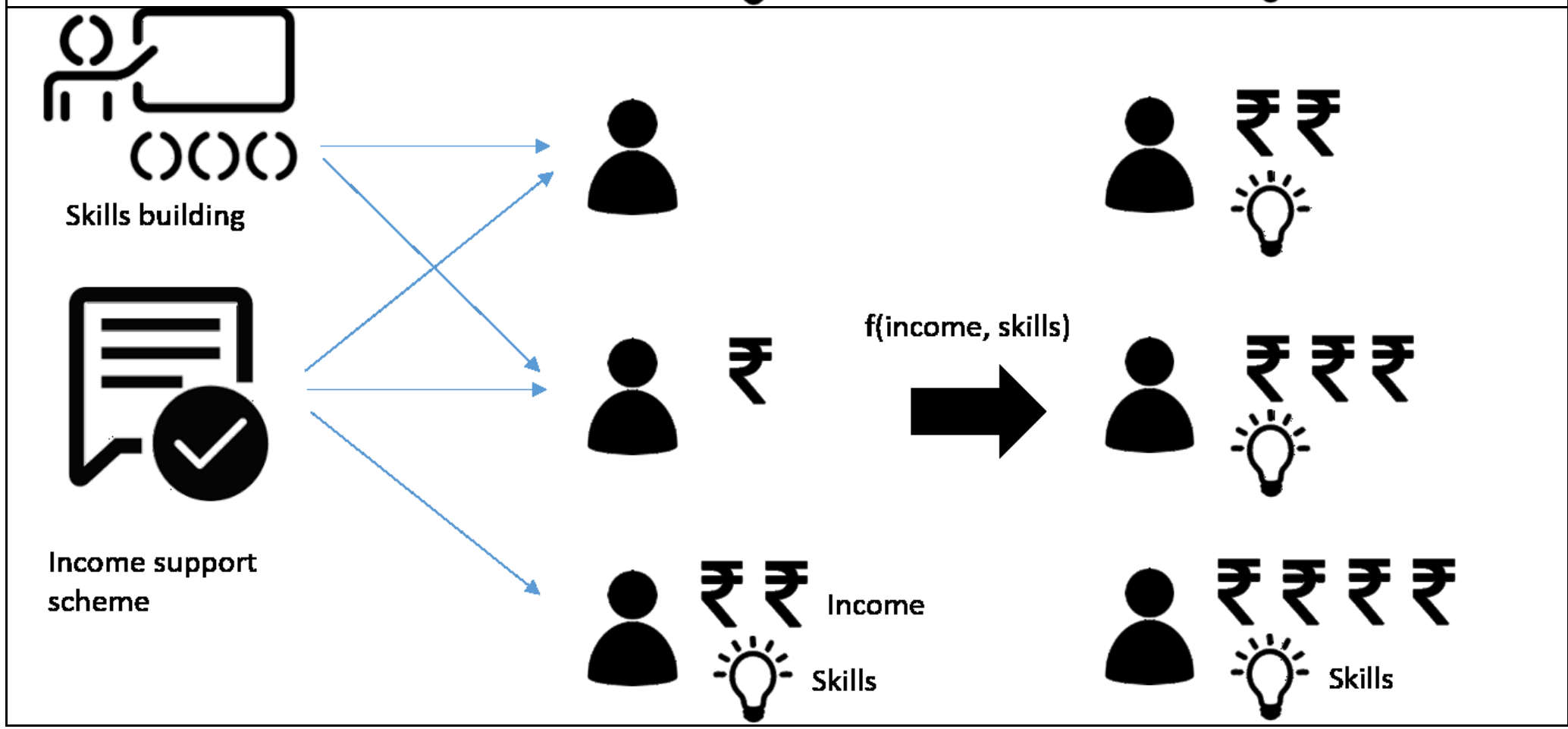

Figure 4b. This augments the setup of Figure 2b with a skills building component which specifically focuses on people with less skill levels. Through this additional training, people are able to do much better in utilizing the income support scheme, leading to a more equitable distribution of resources.

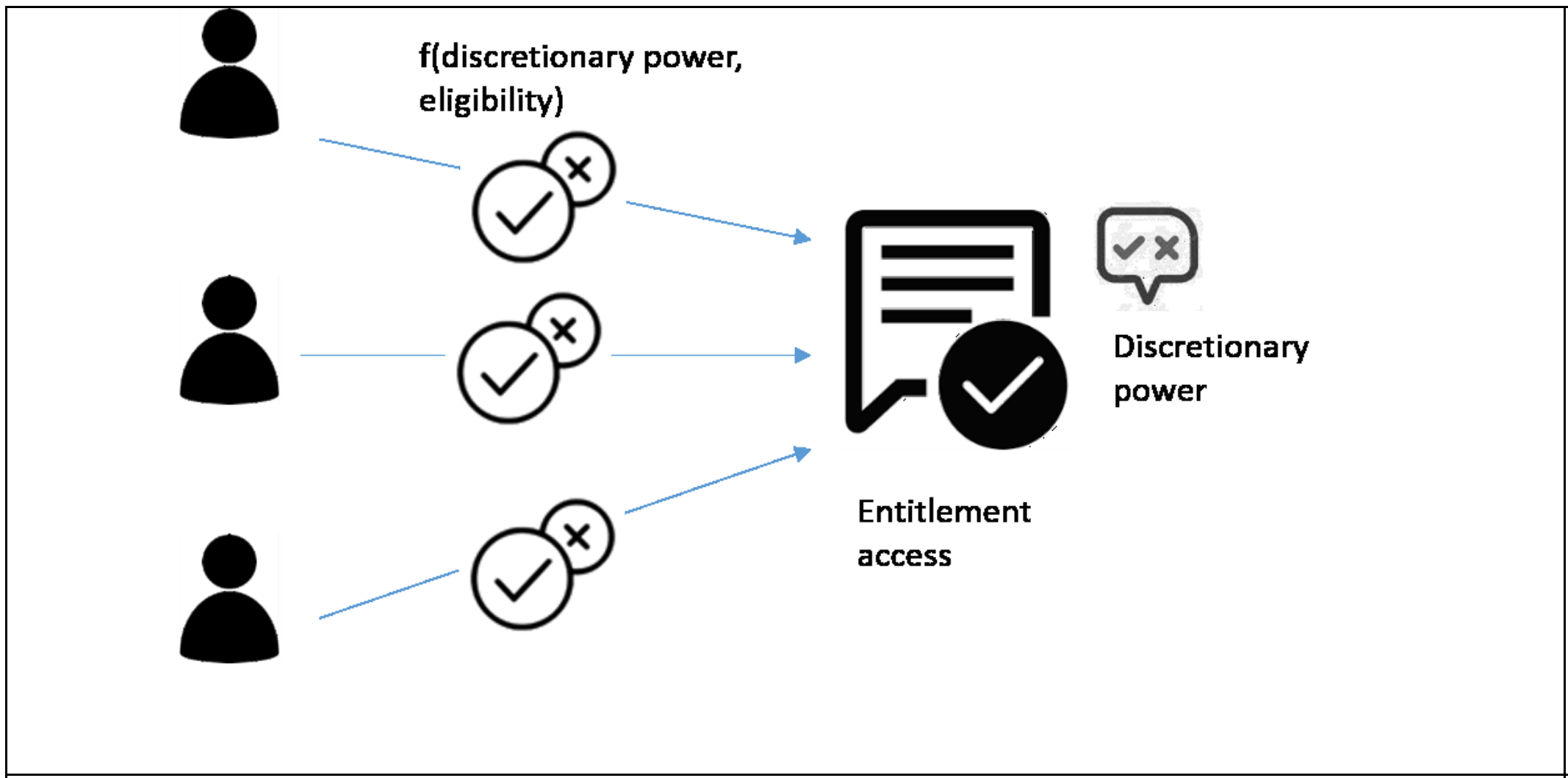

Figure 5a. Discretionary power is a resource in the hands of administrators that can be used to deny entitlements or grievance redressal to people. This is especially true when people do not have access to adequate legal or administrative escalation channels, or a good knowledge of the required processes and documentation. This often leads to bribery as a rent seeking method by the administrators to do the needful about access to entitlements.

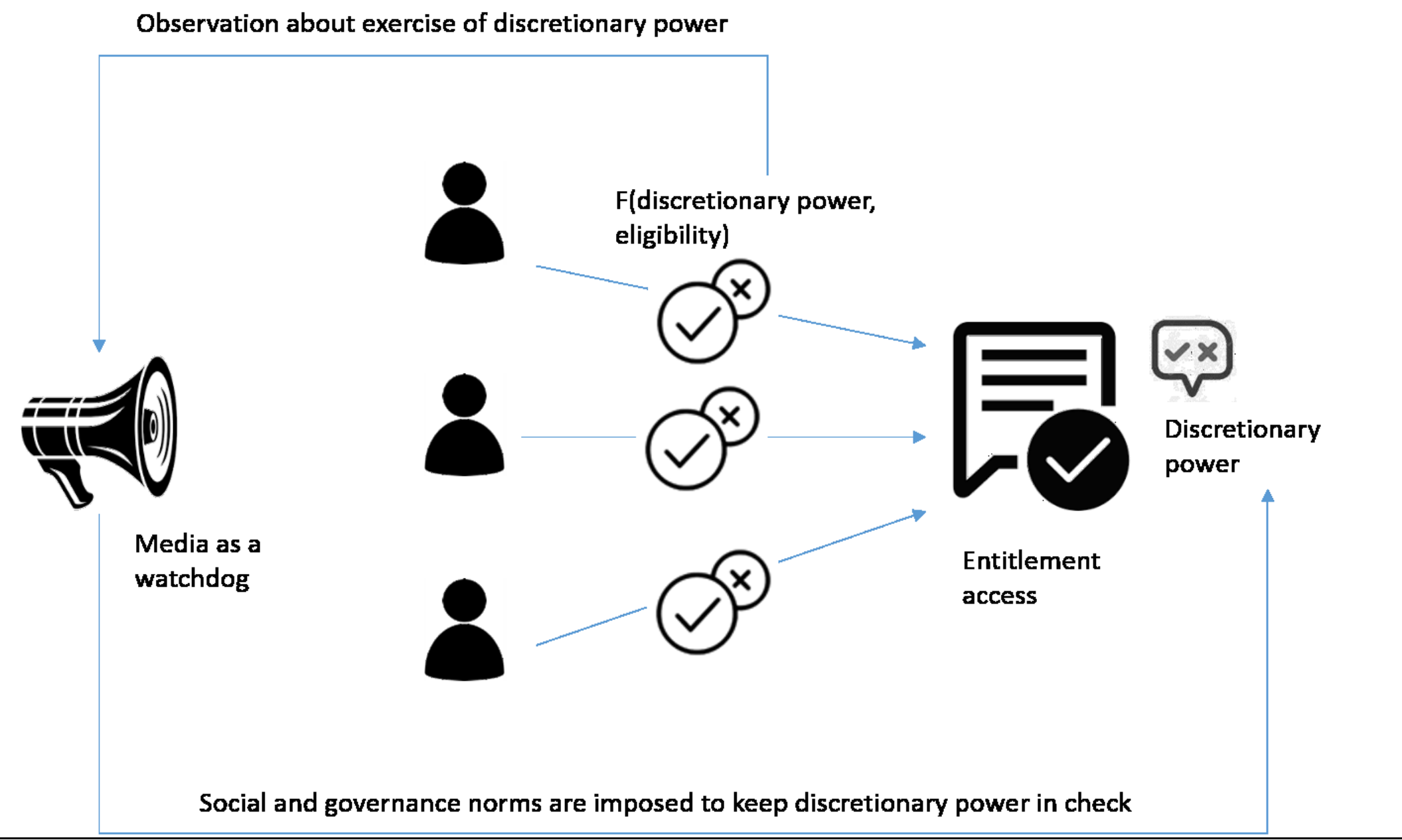

Figure 5b. The media is an important institution to impose checks and balances on power holders. Stories carried in the media about the illegitimate exercise of discretionary power by the administrators, can put pressure on the legal system or higher officials to react and address the rent seeking problems. This feedback function can prevent the discretionary power from increasing indefinitely or being exercised indiscriminately.

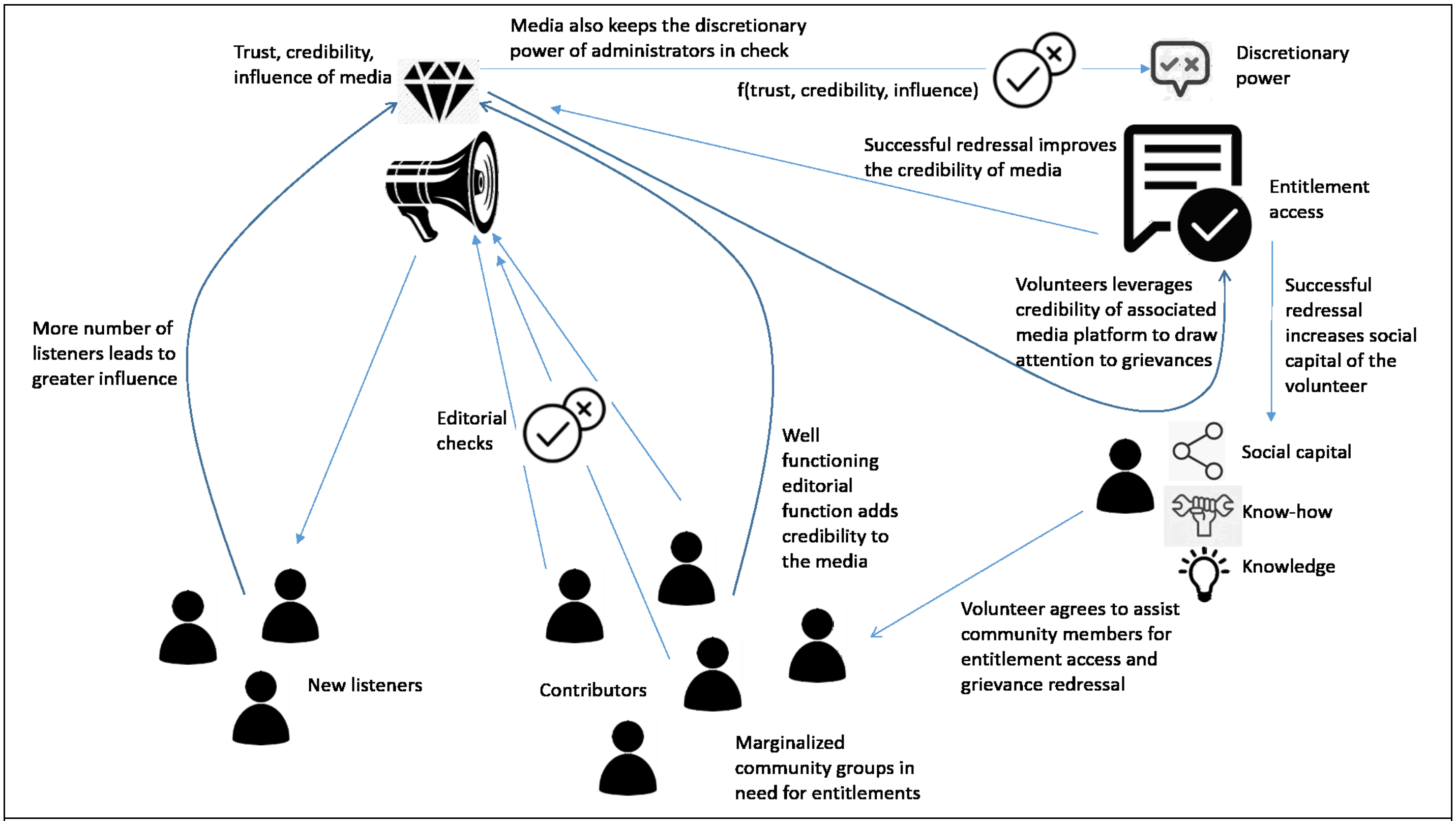

Figure 6. Shown here are three pathways through which the power of a participatory media platform is presumed to increase. Different pathways lead to different types of power as a resource. First, to the left is shown that the larger the audience the media platform has the more influence it would have. Second, towards the middle is shown that demonstrating a well-functioning editorial function will add to the credibility of the media platform. Third, to the right is shown that facilitating grievance redressal as explained earlier in Figure 5b, will also add to the trust placed in the media platform. Similarly, volunteers who facilitate grievance redressal on behalf of marginalized communities will also gain social credibility as a resource. Feedback functions can thus be composed to model these dynamics. For example:

- influence ≈ number_of_users, ie. more the number of users, greater the influential power of the media platform
- credibility ≈ accuracy(user_generated_content_selection), ie. correct decisions about accepting/rejecting user generated content will lead to greater credibility. Note that there may not be any universal notion of correctness of the decisions, rather it may change based on the community priorities of what kind of content they prefer, and would reflect the degree to which the media platform espouses the preferences of the community
- trust += successful_grievance_redressal, ie. with each successful grievance redressal, the trust placed by the community in the media platform will increase

These different forms of power will influence the decision function of whether or not the media platform is able to impose the required checks and balances on the administrators. Such a modeling leads to several questions such as whether all these forms of media power are required, whether credibility or trust is modeled better as a linear function or exponentially increasing function, what is the form of the decision function to bring about action, etc. Reasoning about these dynamics can inform the strategies and priorities of the participatory media platform.